\documentclass[twoside,twocolumn,9pt]{article}
\usepackage{extsizes}
\usepackage[super,sort&compress,comma]{natbib}
\usepackage[version=3]{mhchem}
\usepackage[left=1.5cm, right=1.5cm, top=1.785cm, bottom=2.0cm]{geometry}
\usepackage{balance}
\usepackage{mathptmx}
\usepackage{matlab-prettifier}
\usepackage{booktabs}
\usepackage{sectsty}
\usepackage{graphicx}
\usepackage{lastpage}
\usepackage[format=plain,justification=justified,singlelinecheck=false,font={stretch=1.125,small,sf},labelfont=bf,labelsep=space]{caption}
\usepackage{float}
\usepackage{fancyhdr}
\usepackage{fnpos}
\usepackage[english]{babel}
\usepackage{soul}
\addto{\captionsenglish}{%
  \renewcommand{\refname}{Notes and references}
}
\usepackage{array}
\usepackage{droidsans}
\usepackage{charter}
\usepackage[T1]{fontenc}
\usepackage{setspace}
\usepackage[compact]{titlesec}
\usepackage{hyperref}

\usepackage{epstopdf}

\definecolor{cream}{RGB}{222,217,201}

\begin{document}

\pagestyle{fancy}
\thispagestyle{plain}
\fancypagestyle{plain}{
\renewcommand{\headrulewidth}{0pt}
}

\makeFNbottom
\makeatletter
\renewcommand\LARGE{\@setfontsize\LARGE{15pt}{17}}
\renewcommand\Large{\@setfontsize\Large{12pt}{14}}
\renewcommand\large{\@setfontsize\large{10pt}{12}}
\renewcommand\footnotesize{\@setfontsize\footnotesize{7pt}{10}}
\makeatother

\renewcommand{\thefootnote}{\fnsymbol{footnote}}
\renewcommand\footnoterule{\vspace*{1pt}%
\color{cream}\hrule width 3.5in height 0.4pt \color{black}\vspace*{5pt}}
\setcounter{secnumdepth}{5}

\makeatletter
\renewcommand\@biblabel[1]{#1}
\renewcommand\@makefntext[1]%
{\noindent\makebox[0pt][r]{\@thefnmark\,}#1}
\makeatother
\renewcommand{\figurename}{\small{Fig.}~}
\sectionfont{\sffamily\Large}
\subsectionfont{\normalsize}
\subsubsectionfont{\bf}
\setstretch{1.125} 
\setlength{\skip\footins}{0.8cm}
\setlength{\footnotesep}{0.25cm}
\setlength{\jot}{10pt}
\titlespacing*{\section}{0pt}{4pt}{4pt}
\titlespacing*{\subsection}{0pt}{15pt}{1pt}

\fancyfoot{}
\fancyfoot[LO,RE]{\vspace{-7.1pt}\includegraphics[height=9pt]{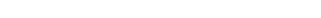}}
\fancyfoot[CO]{\vspace{-7.1pt}\hspace{11.9cm}\includegraphics{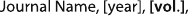}}
\fancyfoot[CE]{\vspace{-7.2pt}\hspace{-13.2cm}\includegraphics{head_foot/RF}}
\fancyfoot[RO]{\footnotesize{\sffamily{1--\pageref{LastPage} ~\textbar  \hspace{2pt}\thepage}}}
\fancyfoot[LE]{\footnotesize{\sffamily{\thepage~\textbar\hspace{4.65cm} 1--\pageref{LastPage}}}}
\fancyhead{}
\renewcommand{\headrulewidth}{0pt}
\renewcommand{\footrulewidth}{0pt}
\setlength{\arrayrulewidth}{1pt}
\setlength{\columnsep}{6.5mm}
\setlength\bibsep{1pt}

\makeatletter
\newlength{\figrulesep}
\setlength{\figrulesep}{0.5\textfloatsep}

\newcommand{\topfigrule}{\vspace*{-1pt}%
\noindent{\color{cream}\rule[-\figrulesep]{\columnwidth}{1.5pt}} }

\newcommand{\botfigrule}{\vspace*{-2pt}%
\noindent{\color{cream}\rule[\figrulesep]{\columnwidth}{1.5pt}} }

\newcommand{\dblfigrule}{\vspace*{-1pt}%
\noindent{\color{cream}\rule[-\figrulesep]{\textwidth}{1.5pt}} }

\makeatother

\twocolumn[
  \begin{@twocolumnfalse}
{\includegraphics[height=30pt]{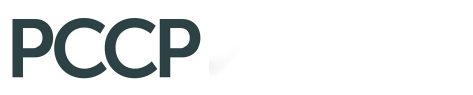}\hfill\raisebox{0pt}[0pt][0pt]{\includegraphics[height=55pt]{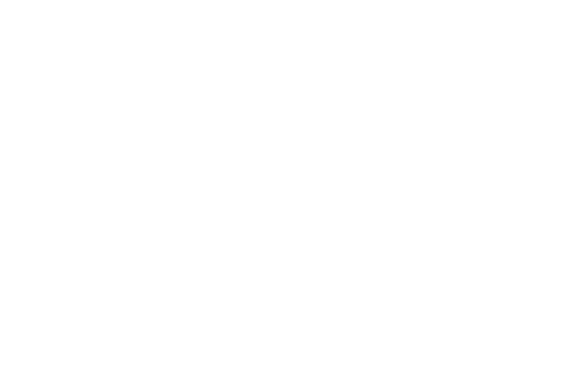}}\\[1ex]
\includegraphics[width=18.5cm]{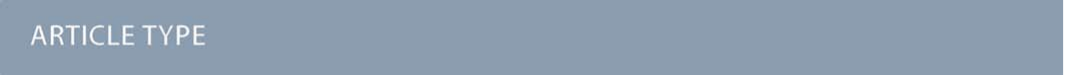}}\par
\vspace{1em}
\sffamily
\begin{tabular}{m{4.5cm} p{13.5cm} }

\includegraphics{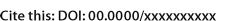} & \noindent\LARGE{\textbf{Singlet Fission in Carotenoid Dimers -- The Role of the Exchange and Dipolar Interactions}} \\
\vspace{0.3cm} & \vspace{0.3cm} \\

 & \noindent\large{Alexandru G.\ Ichert\textit{$^{a,b}$} and William Barford\textit{$^{a,c}$}} \\

\includegraphics{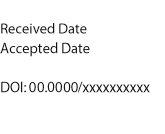} & \noindent\normalsize{A theory of singlet fission in carotenoid dimers is presented which aims to explain the mechanism behind the creation of two uncorrelated triplets. Following the initial photoexcitation of a carotenoid chain to a "bright" $n^1B_u^+$ state, there is ultrafast internal conversion to the intrachain "dark" $1^1B_u^-$ triplet-pair state. This strongly exchanged-coupled state evolves into a pair of triplets on separate chains and spin-decoheres to form a pair of single, unentangled triplets, corresponding to complete singlet fission. The simulated  EPR spectra for lycopene  dimers  shows a distinct spectral signal due to the residual exchange coupling between the triplet-pairs on seperate carotenoid chains.} \\

\end{tabular}

 \end{@twocolumnfalse} \vspace{0.6cm}

  ]

\renewcommand*\rmdefault{bch}\normalfont\upshape
\rmfamily
\section*{}
\vspace{-1cm}


\footnotetext{\textit{$^{a}$~Department of Chemistry, Physical and Theoretical Chemistry Laboratory, University of Oxford, Oxford, OX1 3QZ, United Kingdom.}}
\footnotetext{\textit{$^{b}$~Linacre College, University of Oxford, Oxford, OX1 3JA, United Kingdom. E-mail: alexandru.ichert@chem.ox.ac.uk.}}
\footnotetext{\textit{$^{c}$ ~E-mail: william.barford@chem.ox.ac.uk.}}

\footnotetext{\dag~Supplementary Information available: [1.\ Matrix Representation of the Reduced Two-Triplet Hamiltonian, eqn.~\ref{eq:H_two_triplet}. 2.\ Derivation of the Lindblad Rate Constants.] See DOI: 10.1039/cXCP00000x/}



\section{Introduction}

Singlet fission~\cite{Smith2010,Smith2013} is a photophysical process in which a photoexcited singlet state on one chromophore decays into a pair of correlated electron-hole pairs on separate chromophores, which subsequently dissociates and decoheres into separate triplets. Since the initial discovery of this process in anthracene crystals~\cite{Smith2010,Singh1965}, the mechanism of singlet fission in acenes has been of great interest to researchers because of the potential in overcoming the Shockley-Queisser limit~\cite{Shockley,Hannah_Nozik}. There seems to be a general agreement regarding the mechanism in acenes, with the transition between the photoexcited singlet state to the intermolecular triplet-pair being mediated by virtual charge-transfer states~\cite{Berkelbach2,Aryanpour2015}. In contrast, singlet fission in carotenoids is still poorly understood\cite{Santra2022,BarfordChambers,BarfordLycopene}, perhaps owing to the difference in the electronic states of polyenes compared to acenes.

The electronic states of carotenoids are closely related to those of polyenes. Theoretical calculations~\cite{Valentine2020,Barford2022} have shown that the identity of the $2A_g$--family of "dark" states ($2^1A_g^-$, $1^1B_u^-$, $3^1A_g^-$, ...) is comprised of a linear combination of a strongly-bound singlet triplet-pair and a singlet odd-parity charge-transfer exciton (CTE). In contrast, the optically accessible, or "bright", state $1^1B_u^+$ is predominantly a Frenkel-exciton with some even-parity CTE component. In actuality, due to substituent groups on  carotenoid chains, the electronic states do not possess definite electron-hole symmetry~\cite{WBBook}, with states containing both odd-parity and even-parity CTEs. Manawadu \textit{et al}.~\cite{Manawadu2023} have shown that the odd-parity CTE component of the electronic states gives rise to ultrafast internal conversion between states which have an avoided crossing in the adiabatic representation ($S_1$, $S_2$, ...), or equivalently a crossing in the diabatic representation ($2^1A_g^-$, $1^1B_u^-$, $1^1B_u^+$, ...).

As described by Barford~\cite{BarfordLycopene}, singlet fission in lycopene H-aggregates can be explained in terms of an initial photoexcitation to a "bright" Frenkel-exciton state, which we will call $n^1B_u^+$. In the case where $n=1$, and assuming that the excited chromophore is electronically coupled to at least another lycopene molecule, there are two possibilities:
\begin{enumerate}
    \item that the vertical $2^1A_g^-$ state lies above the vertical $1^1B_u^+$ state, which means that excitation into $1^1B_u^+$ causes a $1^1B_u^+-2^1A_g^-$ level crossing;
    or
    \item that the vertical $2^1A_g^-$ state lies below the vertical $1^1B_u^+$ state, which means that excitation into $1^1B_u^+$ causes a $1^1B_u^+-1^1B_u^-$ level crossing.
\end{enumerate}

\begin{figure}[]
    \centering
    \includegraphics[width=0.75\linewidth]{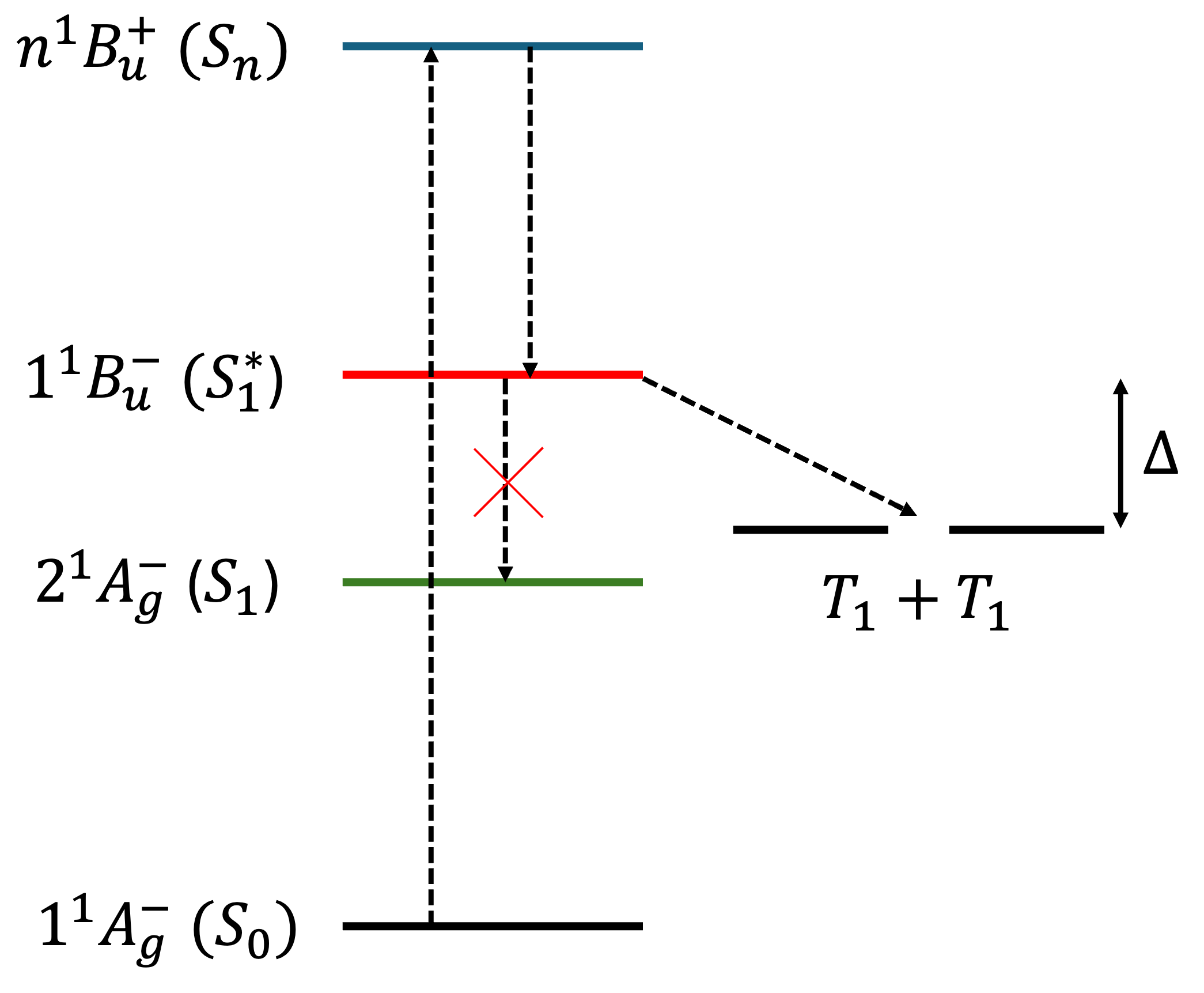}
    \caption{A schematic energy level diagram for some of the electronic states in lycopene\cite{BarfordLycopene} (adiabatic state labels are given in brackets). The singlet, intrachain states are on the left, while the right  the energy level of two interchain separate triplets are shown. Within the Born-Oppenheimer approximation, the interconversion between $1^1B_u^-$ and $2^1A_g^-$ is symmetry forbidden, so we assume bimolecular exothermic singlet fission with an exothermic driving force $\Delta$.}
    \label{fig:ElecStates}
\end{figure}

The first scenario results in ultrafast population transfer from the "bright" state to the "dark" $2^1A_g^-$ state, the lowest energy member of the $2A_g$--family of states. Due to its strongly-bound triplet-pair nature, singlet fission from this state is typically endothermic, and thus unfavorable and unlikely to be the first step in the mechanism. On the other hand, the second scenario would provide a reasonable explanation for the population of the $1^1B_u^-$ state, which lies higher in energy than a pair of uncoupled triplets on separate chromophores and is a prime candidate for the identity of an intermediate in exothermic singlet fission (see Fig.~\ref{fig:ElecStates}). This is inconsistent with the findings of Kundu and Dasgupta~\cite{Kundu2021}, however, where they discovered that low-energy excitation into the $2.5-3.1$ eV range does not result in singlet fission in the aggregate, whilst higher-energy (3.5 eV) photoexcitation does. This has prompted the suggestion that the first scenario is at play\cite{BarfordLycopene}: assuming this, low-energy excitation to the $1^1B_u^+$ state cannot lead to exothermic singlet fission, whilst an excitation to a higher-energy state, $n^1B_u^+$, could populate a higher energy "dark" state in the $2A_g$--family, potentially making singlet fission energetically favourable. In this paper we assume that this "dark" state is $1^1B_u^-$. Once populated, as a consequence of the $\textrm{C}_\textrm{2h}$ symmetry of lycopene, to zeroth order in the Born-Oppenheimer approximation the interconversion from this state to the lower $2^1A_g^-$ state is  symmetry forbidden. However, the interconversion of this strongly-bound triplet-pair to an interchain, spin-entangled triplet-pair is allowed as long as the pair is formed with an overall spin $S=0$. The triplet-pair can then diffuse and spin-decohere to form single, unentangled triplets.

The existence of an intermediate state in the process of singlet fission in carotenoids was postulated based on the work of Kosumi \textit{et al}.~\cite{Kosumi2006}, whose experimental findings uncovered an apparent violation of the energy-gap law. It was suggested that shorter carotenoids relax directly from the "bright" $S_2$ state to $S_1$ via internal conversion, whilst carotenoids which have more than nine conjugated carbon-carbon double bonds relaxed from $S_2$ to $S_1$ via an intermediate state, typically labelled $S^*$. $S^*$ was assumed to provide an alternative relaxation pathway which was faster than the direct $S_2-S_1$ internal conversion. Population of this intermediate state could, in theory, lead to exothermic singlet fission as described above.

Two mechanisms have therefore emerged as popular explanations for the phenomenon, namely one in which the identity of the intermediate is indeed one of the members of the $2A_g$--family of states~\cite{BarfordChambers,BarfordLycopene,Kundu2021,Quaranta2021} and one in which the intermediate presents significant interchain charge-transfer character~\cite{Musser2015,Peng2024}. Here we present a general theory of singlet fission in carotenoid dimers based on the former, i.e., on strongly exchange-coupled triplet-pairs. We apply the model to lycopene dimers in an H-aggregate, with the possibility that the theory can be expanded in the future to cover a larger number of carotenoid chains or even the whole aggregate, providing a foundation for further theoretical developments. We begin by laying out the theoretical background of our model, which is an expanded version of the model initially introduced by Barford and Chambers~\cite{BarfordChambers}. Following that, we discuss the energy spectrum of the system,  the dynamics of singlet fission and our predictions for the EPR signatures of this mechanism. Finally, Appendix A includes a perturbation theory derivation of the eigenstates and spectrum of the lowest nine eigenstates in the limit of exchange coupling and weaker dipolar interactions, whilst Appendix B contains a discussion of how the entanglement entropy vanishes at long times corresponding to complete singlet fission.

\section{Theory of Singlet Fission in Carotenoid Dimers}
\subsection{The triplet-pair basis}

\begin{figure}[t]
    \centering
    \includegraphics[width=1\linewidth]{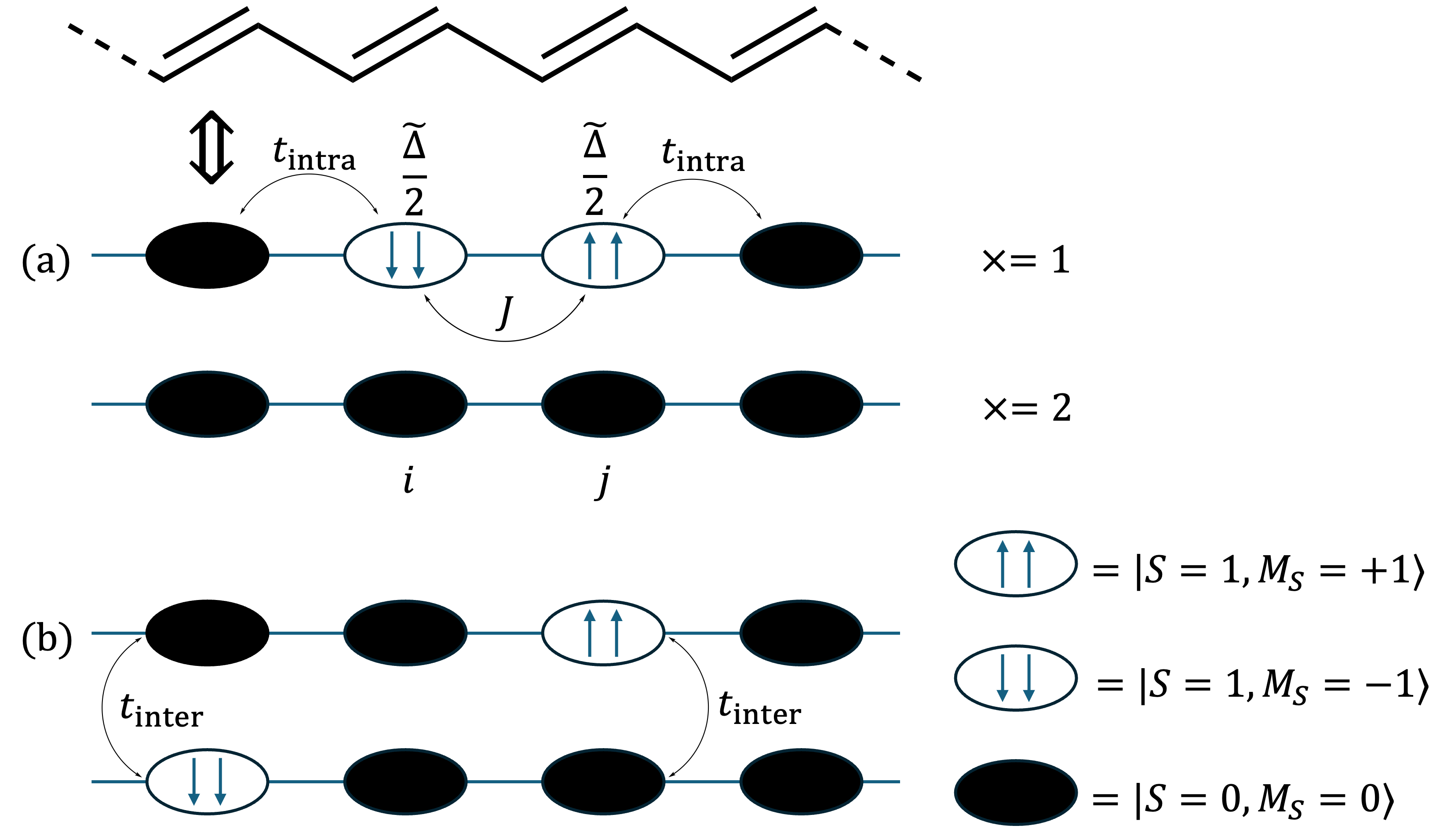}
    \caption{A schematic illustration of the carotenoid dimer. (a) Chain $\times = 1$ shows the intrachain triplet-pair on adjacent C-C dimers \textit{i} and \textit{j}. Each triplet requires $\tilde{\Delta}/2$ energy to be created. The triplet-pair experiences an interaction \textit{J} when occupying neighbouring C-C dimers. Chain $\times = 2$ is in the ground state. (b) the two triplets have dissociated and can be found on two separate chains. $t_\textrm{intra}$ and $t_\textrm{inter}$ are the intrachain and interchain hopping matrix elements respectively. }
    \label{fig:ValenceModel}
\end{figure}

We begin our description of singlet fission in carotenoid dimers by introducing the triplet-pair basis. The model is based on the valence bond theory of strongly correlated systems, with a typical carotenoid dimer illustrated in Fig.~\ref{fig:ValenceModel}. In (a) the chain labelled $\times = 2$ shows the ground state of a single chain, with all C-C dimers in their singlet ground states. A triplet excitation on C-C dimer \textit{i} is represented as $|m_s;~i\rangle$, where $m_s$ can take the values 0 or $\pm 1$. It follows that a pair of triplet excitations on C-C dimers \textit{i} and \textit{j} with spin projection quantum numbers $m_{s,i}$ and $m_{s,j}$ respectively can be described by the tensor product $|m_{s,i};~i\rangle |m_{s,j};~j\rangle$.

Following the initial photoexcitation, we assume that the $n^1B_u^+$ state undergoes ultrafast internal conversion to form a singlet triplet-pair state on a single chain, which we label as $\times = 1$. The initial triplet-pair state, which we take to be the $1^1B_u^-$ state, is
\begin{equation}
{^1}|TT\rangle_{\times=1} \equiv {^1}|\Phi(t=0)\rangle = \sum_{ij\in \times = 1}\Phi_{ij}(t=0){^1}|i,j\rangle,
\end{equation}
where the kets $|i,j\rangle$ form the coupled-spin basis states,
\begin{eqnarray}\label{Eq:2}
\nonumber\\
  ^{1}|i,j\rangle &=& \frac{1}{\sqrt{3}}\left( |+1;i\rangle|-1;j\rangle - |0;i\rangle|0;j\rangle +|-1;i\rangle|+1;j\rangle \right),
\nonumber\\
  ^{3,-1}|i,j\rangle &=& \frac{1}{\sqrt{2}}\left( |0;i\rangle|-1;j\rangle - |-1;i\rangle|0;j\rangle \right),
  \nonumber\\
  ^{3,~0}|i,j\rangle &=& \frac{1}{\sqrt{2}}\left( |+1;i\rangle|-1;j\rangle - |-1;i\rangle|+1;j\rangle \right),
  \nonumber\\
  ^{3,+1}|i,j\rangle &=& \frac{1}{\sqrt{2}}\left( |+1;i\rangle|0;j\rangle - |0;i\rangle|+1;j\rangle \right),
\nonumber\\
  ^{5,-2}|i,j\rangle &=&  |-1;i\rangle|-1;j\rangle,
  \nonumber\\
  ^{5,-1}|i,j\rangle &=& \frac{1}{\sqrt{2}}\left( |-1;i\rangle|0;j\rangle + |0;i\rangle|-1;j\rangle \right),
  \nonumber\\
  ^{5,~0}|i,j\rangle &=& \frac{1}{\sqrt{6}}\left( |+1;i\rangle|-1;j\rangle +
  2|0;i\rangle|0;j\rangle + |-1;i\rangle|+1;j\rangle \right),
  \nonumber\\
  ^{5,+1}|i,j\rangle &=& \frac{1}{\sqrt{2}}\left( |+1;i\rangle|0;j\rangle + |0;i\rangle|+1;j\rangle \right),
  \nonumber\\
  ^{5,+2}|i,j\rangle &=& |+1;i\rangle|+1;j\rangle.
  \nonumber\\
\end{eqnarray}
In general, two triplets can couple to form pairs of overall singlet, triplet or quintet spin multiplicity. We label the triplet and quintet triplet-pairs as $^{2S+1,~M_S}|i,j\rangle$, where $S$ is the overall spin of the coupled pair and $M_S$ is the total spin projection quantum number.

\subsection{The two-chain Hamiltonian}

We write the full, two-chain Hamiltonian as
\begin{equation}\label{eq:H_full}
    \hat{H} = \sum_{\times=1,2}\hat{H}_\textrm{single}^{\times} + \hat{H}_\textrm{double} + \hat{H}_\textrm{inter} + \hat{H}_\textrm{dipolar} + \hat{H}_\textrm{Zeeman}.
\end{equation}
The first term on the right hand side is the \textit{single-chain} Hamiltonian,
\begin{eqnarray}\label{eq:H_single}\nonumber
\hat{H}_{\textrm{single}}^{\times=1,2} &= & \tilde{\Delta}\sum_{i, j>i \in \times}  \left|i,j\rangle \langle i,j\right| + t_{\textrm{intra}}\sum_{i \ne j \in \times}\left(\left|i \pm 1,j\rangle \langle i,j\right| + \textrm{h.c.} \right)\\
 &+&  J\sum_{i \in \times}  \hat{\textbf{S}}_i^{(1)}\cdot\hat{\textbf{S}}_{i+1}^{(1)}.
\end{eqnarray}
The first term on the right hand side of eqn.~\ref{eq:H_single} describes the energy required to excite a pair of triplets on C-C dimers \textit{i} and \textit{j}, where $\tilde{\Delta} = \Delta - \textrm{BE}$. BE is the energy required to dissociate a bound intrachain triplet pair into a pair of noninteracting interchain triplets and $\Delta$ is the exothermic driving energy defined in Fig.~\ref{fig:ElecStates}. The second term defines the hopping of the triplets between adjacent C-C dimers on the same chain, characterised by the matrix element $t_{\textrm{intra}}$. The final term in eqn.~\ref{eq:H_single} describes the spin-dependent interaction between two intrachain triplets on adjacent C-C dimers. It was shown~\cite{Barford2022} that triplets occupying neighbouring C-C dimers experience an enhanced interaction due to the hybridisation between the singlet triplet-pair and singlet odd-parity CTE components of the "dark" $2A_g$--family of states. We write this overall spin-dependent interaction as
\begin{equation}\label{eq:exchange}
   J \hat{\textbf{S}}_{i}^{(1)}\cdot \hat{\textbf{S}}_{i+1}^{(1)},
\end{equation}
where $\hat{\textbf{S}}_{i}^{(1)}$ is the spin-1 operator
acting on C-C dimer \textit{i} and \textit{J} is the exchange parameter describing the inter-triplet interaction. It follows that a singlet triplet-pair experiences an attraction $2J$, a triplet triplet-pair experiences an attraction $J$ and a quintet triplet-pair experiences a repulsion $J$~\cite{Kollmar,Barford2022,BarfordChambers}.

$\hat{H}_\textrm{double}$ is the Hamiltonian for a pair of triplets on separate chains,
\begin{eqnarray}\label{eq:H_double}
\nonumber
\hat{H}_\textrm{double} & = & t_{\textrm{intra}}\sum_{i \in \times=1} \sum_{j \in \times=2} \left(\left|i \pm 1,j\rangle \langle i,j\right|+ \textrm{h.c.} \right)\\
& + &\left(\left|i,j \pm 1\rangle \langle i,j\right| + \textrm{h.c.} \right),
\end{eqnarray}
used to describe the hopping motion of triplets between neighouring C-C dimers on the same chain.

The third term in eqn.~\ref{eq:H_full} is the \textit{interchain} Hamiltonian,
\begin{eqnarray}\label{eq:H_inter}
\nonumber
 \hat{H}_\textrm{inter} = t_{\textrm{inter}} \sum_{\times = 1}^{2} \sum_{i_{\times}} \sum_{j_{\times}>i_{\times}}\left(\left|i_{\times},j_{\times}\rangle \langle i_{\times},j_{\bar{\times}}\right|+ \textrm{h.c.}\right)\\
  +\left(\left|i_{\times},j_{\times}\rangle \langle i_{\bar{\times}},j_{\times}\right|+ \textrm{h.c.}\right),
\end{eqnarray}
where $i_{\times}$ and $i_{\bar{\times}}$ mean the \textit{i}-th C-C dimer on opposite chains. This Hamiltonian describes the hopping of individual triplets between adjacent C-C dimers on separate chains, as shown in Fig.~\ref{fig:ValenceModel}. The interchain hopping is described by the matrix element $t_{\textrm{inter}}$, which is also used to quantify the electronic coupling between the carotenoid chains.

In addition to the interactions described so far, we also include the \textit{intratriplet} dipolar interaction,
\begin{equation}\label{eq:H_dipolar}
\hat{H}_{\textrm{dipolar}} = \sum_i {\hat{H}_{\textrm{dipolar}}}^i,
\end{equation}
with
\begin{eqnarray}\label{eq:H_dipolar_a}\nonumber
{\hat{H}_{\textrm{dipolar}}}^i &=& {\hat{\textbf{S}}_i^{(1)}} \cdot \textbf{D} \cdot \hat{{\textbf{S}}}_{i}^{(1)}
\\
\nonumber
&=& D\left((\hat{S}_{iz}^{(1)})^2 - \frac{1}{3}(\hat{S}_i^{(1)})^2\right) + \frac{E}{2}\left((\hat{S}_{i+}^{(1)})^2 + (\hat{S}_{i-}^{(1)})^2\right)\\
\end{eqnarray}
and zero-field splitting (ZFS) parameters~\cite{WeilBook} defined as
\begin{eqnarray}
&&D = \frac{3}{2}D_z,\\
&&E = \frac{1}{2}(D_x - D_y),\\
&&D_x + D_y + D_z = 0,\\
&&|D_z|\geq|D_x|\geq|D_y|\implies|D|\geq 3|E|.
\end{eqnarray}
We assume that the principal-axis system (i.e., the axis system which diagonalises the dipolar coupling matrix \textbf{D}) coincides with the axes of symmetry of the molecule. Furthermore, we assume that the two triplets are always colinear. Therefore, in a perfect H-aggregate, $\hat{H}_{\textrm{dipolar}}$ possesses permutation symmetry and thus preserves the spin-parity of the triplet-pair states -- it can only mix the singlet and quintet spin-subspaces. (However, in the case of misaligned chains, the symmetry is broken and singlet-triplet-quintet mixing can occur~\cite{Collins2019,Chen2019,Burdett2013,Tapping2016}.)

The final interaction we consider is the Zeeman interaction between an external applied magnetic field and particles of spin-1,
\begin{eqnarray}
{\hat{H}_{\textrm{Zeeman}}}&=& \sum_i \mu_B\textbf{B}\cdot\textbf{g}\cdot\hat{{\textbf{S}}}_{i}^{(1)}.
\end{eqnarray}
The magnetic field plays the important role of removing the arbitrary choice of the axis of spin quantisation. Whenever there is an applied magnetic field, the spin projection quantum number is only defined parallel to the direction of the field (i.e., only the eigenstates of the Zeeman Hamiltonian have a defined $M_S$). Otherwise, the conventional choice for spin quantisation is the principal-axis \textbf{Z}.

\subsection{Time evolution - the quantum Liouville equation}

The dynamics of the system are exactly described by the time-dependent density operator, $\hat{\rho}(t)$. The time evolution of the density operator is given by the quantum Liouville-von Neumann equation,
\begin{equation}\label{eq:QLE}
\frac{d\hat{\rho}}{dt} = -\frac{i}{\hbar}\left[ \hat{H}_{\textrm{system}},\hat{\rho} \right] + \hat{\hat{D}}\hat{\rho},
\end{equation}
where the first term on the right hand side describes the unitary evolution of the system in the absence of system-bath interactions (with eqn.~\ref{eq:H_full} replacing $\hat{H}_{\textrm{system}}$), whilst the second term is a dissipator used to describe the effect of system-bath interactions on the system. We consider three contributions to the dissipator, one arising from the nonmagnetic, spin-conserving interactions and two arising from magnetic, spin-nonconserving interactions.

Assuming the secular approximation, which explicitly decouples the evolution of the populations from those of the coherences, and defining $\omega_{ab} = (E_a-E_b)/\hbar \geq 0$, eqn.~\ref{eq:QLE} can be re-written as
\begin{equation}\label{eq:QLE_pops}
    \frac{dP_a}{dt}=-\sum_{b\neq a} \left( k_{ab}P_a - k_{ba}P_b \right),
\end{equation}
where $P_a\equiv\rho_{aa}$ is the population of the Hamiltonian eigenstate $|\Psi_a\rangle$ and $k_{ab}$ is the rate of population transfer from eigenstate $a$ to eigenstate $b$. The equivalent equation for the coherences is
\begin{equation}
    \frac{d\rho_{ab}}{dt}=-i\omega_{ab}\rho_{ab}-2\Gamma_{ab}(1-\delta_{ab})\rho_{ab},
\end{equation}
where $\Gamma_{ab} = \frac{1}{2}(\gamma_{a}+\gamma_{b})$ and $\gamma_a = \frac{1}{2}\sum_{b\neq a}k_{ab}$.

The rates $\{k_{ab}\}$ are written in terms of spin-conserving and spin-nonconserving contributions,
\begin{equation}
    k_{ab} = k_{ab}^{SC}+k_{ab}^{SNC}.
\end{equation}
The spin-conserving dynamics are accounted for by the Redfield tensor and $\{k_{ab}^{SC}\}$ are given by
\begin{eqnarray}
&k_{ab}^{SC} =& \left(\frac{2\lambda}{\hbar}\right)J(\omega_{ab})(n(\omega_{ab})+1)C_{ab}^{SC},\\
&k_{ba}^{SC} =& \left(\frac{2\lambda}{\hbar}\right)J(\omega_{ab})n(\omega_{ab})C_{ab}^{SC},
\end{eqnarray}
where $n(\omega) = (e^{\omega/k_BT}-1)^{-1}$ is the Bose distribution function, $J(\omega) = \omega\omega_0/(\omega^2 + \omega_0)$ is the Debye spectral function, and $\lambda$ is the bath reorganisation energy, which quantifies the strength of the system-bath interactions. The eigenfunctions need to spatially overlap in order to transfer population, so the spin-conserving overlap factor is defined as
\begin{equation}
    C_{ab}^{SC} = 2 \sum_{m} \left| S_{ma} \right|^{2} \left| S_{mb} \right|^{2},
\end{equation}
where \textbf{S} is the matrix whose columns are the eigenvectors of the full Hamiltonian in the coupled-spin basis.

The spin-nonconserving rates have been derived by calculating the effect of two Lindblad (spin-1/2) operators acting on the spin-coupled basis set. We take $\hat{L}_1=\hat{S}_x/\hbar = \left( \hat{S}_+ + \hat{S}_- \right)/2\hbar$ and $\hat{L}_2=\hat{S}_z/\hbar$ to represent the effects of longitudinal ($T_1$) and transverse ($T_2$) spin-dephasing, respectively. The rate of spin-nonconserving population transfer between two eigenstates $a$ and $b$ is
\begin{equation}\label{eq:r_p}
    k_{ab}^{SNC}=\sum_{p=1}^{18}r_p C^{SNC}_{ab,p},
\end{equation}
where the sum is over the index $p$ which defines a pair of spin-coupled states which are able to transfer population (see the Supplementary Information for more details on the relative rates, $r_p$). The spin-nonconserving overlap factor is defined as
\begin{equation}
    C_{ab,p}^{SNC} = \sum_{m,\bar{m}\in p} \left| S_{ma} \right|^{2}\left| S_{\bar{m}b} \right|^{2} + \left| S_{mb} \right|^{2}\left| S_{\bar{m}a} \right|^{2},
\end{equation}
where $m$ and $\bar{m}$ label the spin-coupled basis states belonging to index $p$ such that both basis states correspond to the same $i$, $j$ and $\times$ configuration, but different $S$ and $M_S$. It is trivial to show that the spin-conserving rates follow detailed balance and we explicitly enforce that the spin-nonconserving rates obey
\begin{equation}
    k_{ba}^{SNC} = k_{ab}^{SNC}e^{-\omega_{ab}\hbar/k_BT},
\end{equation}
meaning that the overall rates follow detailed balance. Thus, we ensure that the final populations are the same as given by the Boltzmann distribution.

Eqn.~\ref{eq:QLE_pops} can be solved by casting it into the matrix form,
\begin{equation}
    \frac{dP_a}{dt}=\sum_{\forall b}K_{ab}P_b,
\end{equation}
which yields the solutions
\begin{equation}
    P_a(t)=\sum_{bc}L_{ab}~e^{\lambda_bt}~L^{-1}_{bc}P_c(0).
\end{equation}
We define \textbf{K} as the matrix of rate constants, \textbf{L} as the matrix whose columns are eigenvectors of \textbf{K} and $\{\lambda\}$ as the corresponding eigenvalues. Finally, $P_c(0)$ is the initial condition,
\begin{equation}
    \hat{\rho}(t=0) = |1^1B_u^-\rangle\langle 1^1B_u^-|.
\end{equation}

\subsection{Parameters}
A list of the  parameters relevant for lycopene dimers~\cite{BarfordLycopene} used in this paper is given in Table~\ref{tab:parameters}.

\begin{table}[h]
    \centering
    \begin{tabular}{cc}
        \toprule
        \textbf{Parameter} & \textbf{Value} \\
        \midrule
        Intrachain triplet hopping integral, $t_{\textrm{intra}}$ & 0.88 eV  \\
        Interchain triplet hopping integral, $t_{\textrm{inter}}$ & 0.0088 eV\\
        Intrachain triplet exchange interaction, $J$ & 1.23 eV  \\
        Exothermic driving energy, $\Delta$ & 0.32 eV  \\
        ZFS parameter, $D$ & $0.01$ meV  \\
        ZFS parameter, $E$ & $0.001$ meV  \\
        Reorganisation energy, $\lambda$ & 0.05 eV  \\
        Spectral function cut-off frequency, $\omega_0$ & 0.20 eV  \\
        Longitudinal magnetic dephasing time, $T_1$ & 5 ns  \\
        Transverse magnetic dephasing time, $T_2$ & 10 ns  \\
        EPR spectrometer frequency, $\nu$ & 9.5 GHz \\
        Temperature, $T$ ($k_BT$) & 300 K (26 meV) \\
        Derived value of $J_1$ & 0.05 meV \\
        Derived value of $J_2$ & 0.12 meV \\

        \bottomrule
    \end{tabular}
    \caption{Values of parameters used in this work. (See ref.~\cite{BarfordLycopene} for more details.)}
    \label{tab:parameters}
\end{table}

\section{Results}
\subsection{The two-chain spectrum}

\begin{figure}[t]
    \centering
    \includegraphics[width=1\linewidth]{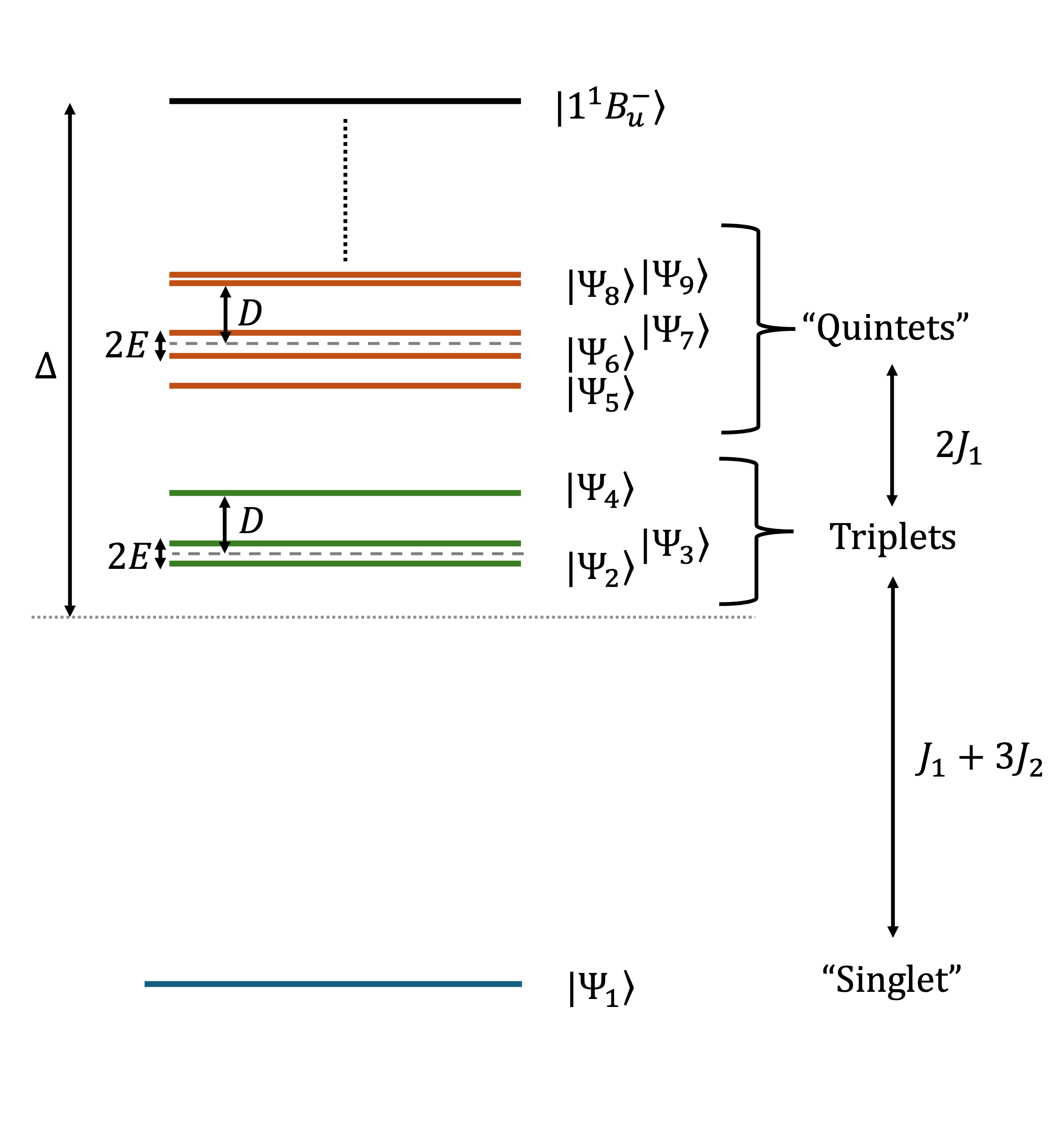}
    \caption{Schematic illustration of the low-energy  zero-field spectrum for  exchange-coupled (predominately interchain) triplet-pairs in lycopene dimers. The eigenstates are labelled according to the spin character which best defines them, with the "singlet" being the eigenstates with mainly singlet character and likewise for the "quintets". The triplets and two out of the five quintet states are pure spin eigenstates. Whilst the triplets only couple with other triplet states, the extent of dipolar coupling between the singlet and quintets is strongly dependent on the energy separation between the spin-subspaces. Also shown is the exothermic driving energy $\Delta$, which is the energy difference between the initial intrachain state $|1^1B_u^-\rangle$ and the barycentre of the thermally accessible, lowest nine eigenstates. In this paper the exchange parameters defined in eqn.~\ref{eq:H_two_triplet}, $J_1$ and $J_2$, take the values $0.05$ and $0.12$ meV, respectively. $k_BT\gg \Delta E_{9-1}$, whereas the energy difference between the tenth and first lowest eigenvalues, $\Delta E_{10-1} \approx 176$ meV $\gg k_BT$. The eigenstates and eigenvalues determined perturbatively are given in Appendix A.}
    \label{fig:ZF_Spec}
\end{figure}

For our choice of parameters (i.e., strongly exchanged-coupled triplet-pairs in the exothermic singlet fission regime), the eigenstates of the total Hamiltonian, eqn.~\ref{eq:H_full}, will take  the general form
\begin{equation}
    |\Psi_{i}\rangle = \frac{a_i}{\sqrt{2}}\left(|TT\rangle_1 + |TT\rangle_2\right) +b_i|T...T\rangle_{1-2},
\end{equation}
with the low-energy zero-field spectrum shown in Fig.~\ref{fig:ZF_Spec}. Coupling between the chains via $\hat{H}_\textrm{inter}$ causes the hybridisation between the intrachain states $|TT\rangle_{\times=1,2}$ and interchain states $|T...T\rangle_{1-2}$, which has been previously explored by Barford and Chambers~\cite{BarfordChambers}. Briefly, strong interchain coupling causes the intrachain character of the (mostly interchain) eigenstates shown in Fig.~\ref{fig:ZF_Spec} to increase. We model singlet fission as strongly exothermic, i.e., $|a_i|^2\ll|b_i|^2$.

A useful quantity initially defined by Barford and Chambers~\cite{BarfordChambers} is $\Delta E_{9-1}$, the energy difference between the thermally accessible eigenstates $|\Psi_9\rangle$ and $|\Psi_1\rangle$. If $\Delta E_{9-1}\ll k_BT$ then the population of the initial state $|1^1B_u^-\rangle$ becomes fully equilibrated between the nine lowest eigenstates, with higher energy states being separated by much larger energy gaps and are generally not thermally accessible (the energy difference between the lowest eigenvalue and the tenth lowest eigenvalue, $\Delta E_{10-1}$, is ca.~176 meV $\gg k_BT$). As these nine eigenstates correspond to interchain, hybridised coupled-spin states, the equal population of these levels corresponds to complete singlet fission.

We can understand this by introducing a two-triplet reduced model exclusively describing these nine eigenstates, namely
\begin{eqnarray}\label{eq:H_two_triplet} \nonumber
    \hat{H}_\textrm{reduced} & =& J_1\hat{\textbf{S}}_{A}^{(1)}\cdot \hat{\textbf{S}}_{B}^{(1)} - J_2\left(\hat{\textbf{S}}_{A}^{(1)}\cdot \hat{\textbf{S}}_{B}^{(1)}\right)^2    \\
     &+& \hat{H}_\textrm{dipolar}+ \hat{H}_\textrm{Zeeman},
\end{eqnarray}
where $A$ and $B$ label the two triplets.
The exchange term, with  parameter $J_1$, is a residual interaction caused by the virtual occupation  of  intrachain states (which experience the strong exchange coupling, eqn.~\ref{eq:exchange}).
The biquadratic exchange term, with  parameter $J_2$, is required to model the delocalization of each triplet on its chain. eqn.~\ref{eq:H_two_triplet} precisely reproduces the spectrum shown in Fig.~\ref{fig:ZF_Spec}.  The triplet and quintet manifolds are separated by $0.1$ meV (i.e., $J_1 = 0.05$ meV), while the singlet state is found $0.4$ meV below the triplet manifold (i.e., $J_2 = 0.12$ meV). This means that $\Delta E_{9-1} \approx 0.51$ meV, much smaller than the thermal energy $k_BT = 26$ meV at $300$ K.

\begin{figure}[t!]
    \centering
    \includegraphics[width=1\linewidth]{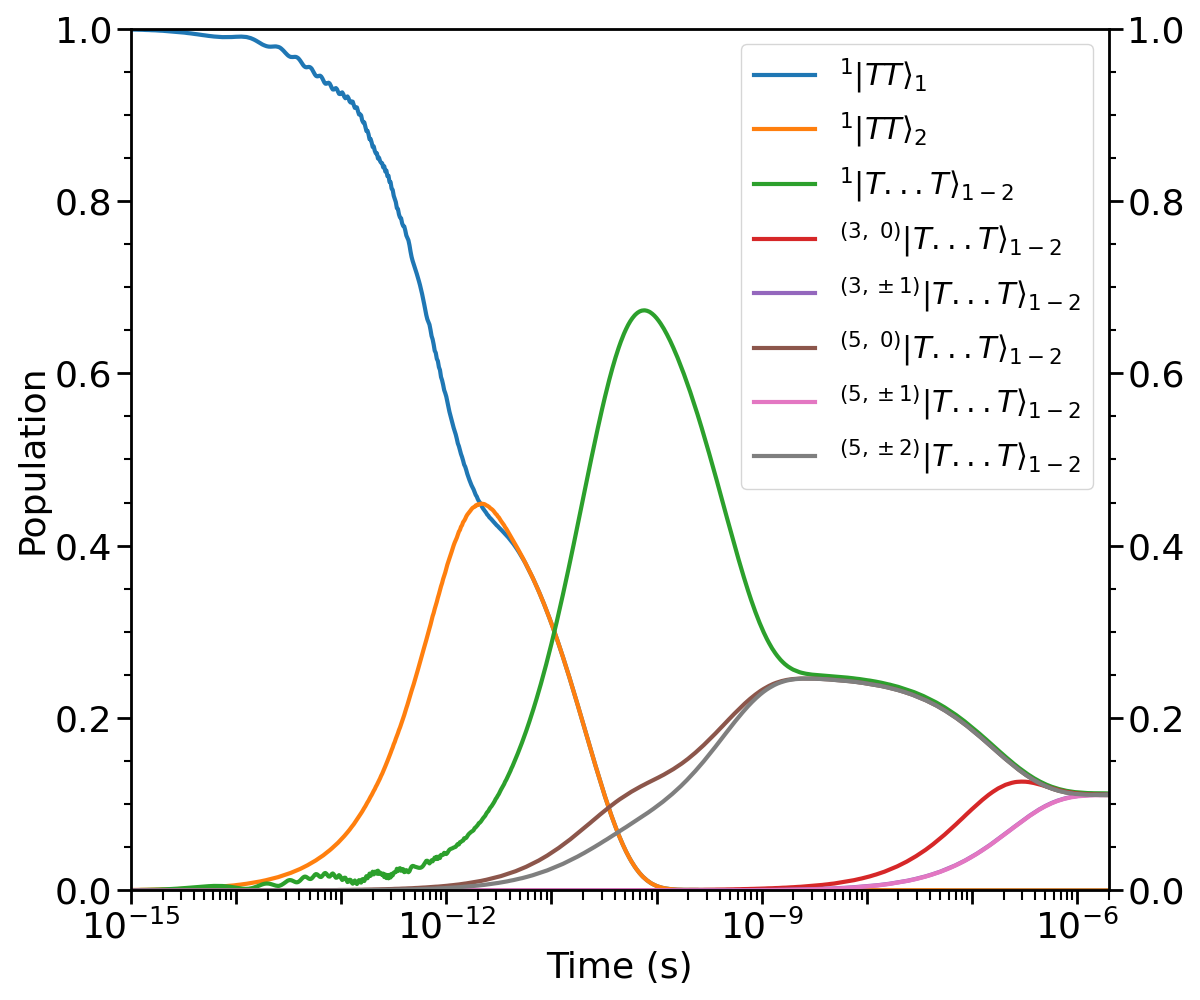}
    \caption{Populations as a function of time of the intrachain states, ${^1}|TT\rangle_{1}$ and ${^1}|TT\rangle_{2}$, and interchain states, ${^{(2S+1,M_S)}}|T...T\rangle_{1-2}$. The dynamics can be divided into four time regimes, which are described in the main text. States of same spin and magnitude of $M_S$ are populated simultaneously. Note that the rate of population of the triplet and quintet states with $M_S=\pm1$ are similar enough that the two lines corresponding to those populations (purple and pink) overlap.}
    \label{fig:PopDyn_ZF}
\end{figure}

The spin character of the eigenstates is determined by the complex interplay between the anisotropic dipolar interaction and the exchange and Zeeman interactions. In the strong-exchange limit,  total spin is a good quantum number. However, formally only the triplets and three of the quintets have definite spin, whilst two of the quintets mix with the singlet to form singlet-quintet states. This is further explained in Appendix A by using a perturbative approach to calculate the eigenvalues and eigenstates of $\hat{H}_\textrm{reduced}$. At high-field all eigenstates have definite spin because of the dominating Zeeman interaction.

Within the reduced two-triplet model, the equal population of the eigenstates means that the density matrix can be written as
\begin{equation}
    \hat{\rho} = \frac{1}{9}\sum_{i=1}^{9}|\Psi_i\rangle\langle\Psi_i|=\frac{1}{9}\sum_{S,M_S}|S,M_S\rangle\langle S,M_S|.
\end{equation}
Equivalently, a rotation of this density matrix to the spin-uncoupled basis $\{ |s_A=1,m_{s,A}\rangle|s_B=1,m_{s,B}\rangle \}$ explicitly shows the spin-decoupling of the triplets. In other words, thermalization has  mixed the eigenstates of eqn.~\ref{eq:H_two_triplet} to form spin-uncorrelated, exchange-coupled single triplets on separate chains. In this limit, $\langle S^2 \rangle = 4 \hbar^2$ and, as shown in Appendix B, the thermalisation of the triplet-pair is accompanied by a vanishing of the entanglement entropy.



\subsection{Dynamical Simulations}

Previous work by Barford~\cite{BarfordLycopene} investigated the dynamics involved in singlet fission whilst assuming only transverse spin-dephasing in an axially symmetric system (i.e., $E = 0$). Here we assume that the system is of orthorhombic symmetry (meaning that the dipolar interaction will hybridise the singlet with the $M_S=0$ quintet as well as quintets of $M_S=\pm2$ spin-projection). This assumption, together with the inclusion of longitudinal spin-dephasing, required us to expand the basis set to include the full nine-spin state space spanned by $|T\rangle\bigotimes|T\rangle$, listed in eqn.\ \ref{Eq:2}

The results of the zero-field dynamical simulation are shown in Fig.~\ref{fig:PopDyn_ZF}. As previously mentioned, we take our initial state to be $|1^1B_u^-\rangle\equiv~^{1}|TT\rangle_1$. Within ca.~2 ps we observe weak coherent population transfer between the intrachain states via the interchain state $^1|T...T\rangle_{1-2}$. After this, we notice that the intrachain population is fully transferred to the interchain states at ca.~100 ps due to the spin-conserving system-bath interactions. This corresponds nicely to the first step in the frequently used Merrifield quantum-kinetic mechanism~\cite{Merrifield,Weiss2017,Tayebjee2017},
\begin{equation}
   ^{1}|TT\rangle_{1} \rightarrow {^1}|T...T\rangle_{1-2},
\end{equation}
with the caveat that we take the bound triplet-pair state to be our starting condition and not the $^{1}|S_1S_0\rangle$ state.

The dipolar interaction, eqn.~\ref{eq:H_dipolar}, mixes the singlet and quintet states such that spin-conserving population transfer results in the gradual population of the $^{5,0}|T...T\rangle_{1-2}$ and $^{5,\pm2}|T...T\rangle_{1-2}$ states. The relative rates of population of the quintet states strongly depend on the magnitude of the ZFS parameters which mix the spin-states. In our case, the ZFS parameter $D$ couples the singlet and the $^{5,0}|T...T\rangle_{1-2}$ whereas $^{5,\pm2}|T...T\rangle_{1-2}$ are coupled to the singlet state by the $E$ parameter. We take $D=10E$, so we indeed expect that the rate of population of $^{5,0}|T...T\rangle_{1-2}$ exceeds that of $^{5,\pm2}|T...T\rangle_{1-2}$.

\begin{figure}[t]
    \centering
    \includegraphics[width=1\linewidth]{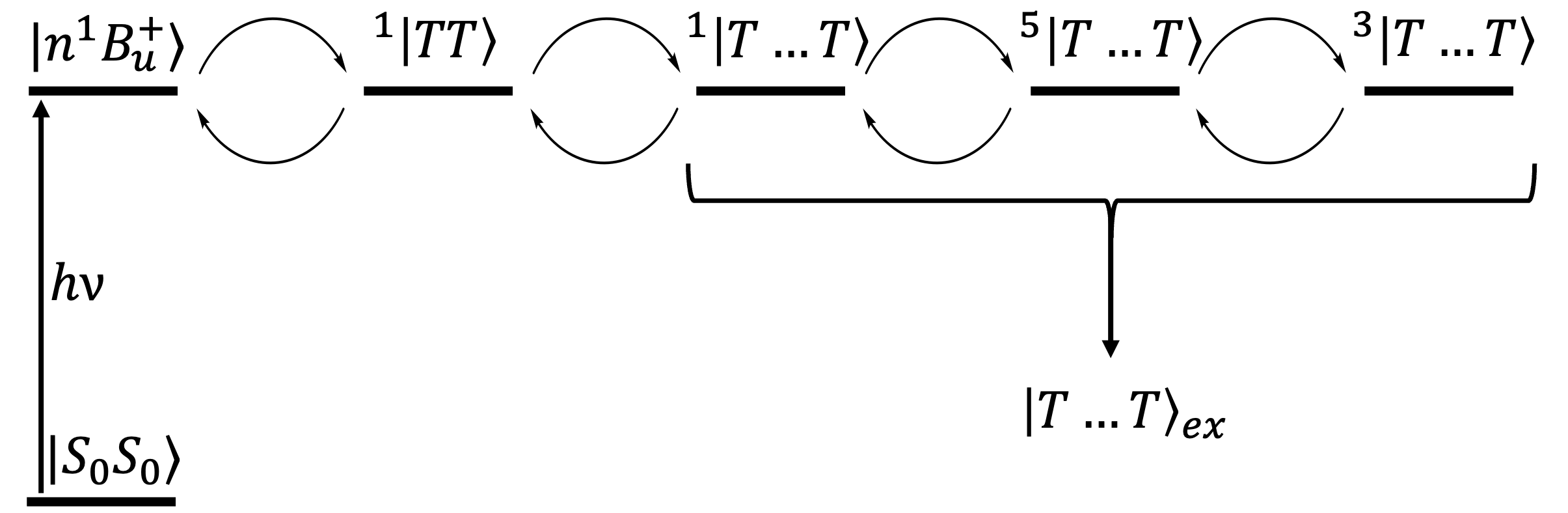}
    \caption{The mechanism of singlet fission in carotenoid dimers proposed in this paper. The initially photoexcited state $|n^1B_u^+\rangle$ undergoes ultrafast internal conversion to a "dark", intrachain triplet-pair state. This state then evolves into a singlet, interchain state $^1|T...T\rangle$ which is mixed with the quintet states by the dipolar interaction. Subsequent spin relaxation populates the quintet and triplet sublevels, which are then thermally mixed with the singlet to form the separate, unentangled 
    triplets $|T...T\rangle_{ex}$.}
    \label{fig:mechanism}
\end{figure}

From ca.~100 ps to ca.~2 ns the spin-conserving population transfer begins to equilibrate the population of the initial state between the singlet and three quintets. The dipolar interaction will affect the population dynamics depending on the orientation of the axis of spin-quantisation due to the orientation-dependent hybridisation of the spin-coupled basis states. However, due to the nature of the assumed symmetry of our system, $\hat{H}_\textrm{dipolar}$ is unable to cause mixing between the $^{5,\pm1}|T...T\rangle_{1-2}$ states and the other quintets and the singlet. Therefore, whatever the orientation of the axis of spin-quantisation (i.e., that of an applied magnetic field), within ca.~2 ns the population of the initial state will be distributed across the same four spin-coupled interchain levels as in Fig.~\ref{fig:PopDyn_ZF}.

Finally, after ca.~2 ns and before ca.~2 $\mu$s the spin-nonconserving population transfer arising from the transverse and longitudinal spin-dephasing equilibrates the total population across the nine lowest interchain states, illustrated in Fig.\ \ref{fig:ZF_Spec}. We observe that the population of each state reaches a thermal equilibrium value of approximately $\frac{1}{9}$ at ca.~2 $\mu$s, meaning that spin-conserving and spin-nonconserving thermalization processes are able to effectively mix the eigenstates in order to form the separate, spin-uncorrelated triplets. In Appendix B we show that this equilibrium population coincides with $\langle S^2 \rangle = 4\hbar^2$ and zero entanglement entropy. This corresponds to complete singlet fission, for which the proposed mechanism is shown in Fig.~\ref{fig:mechanism}.

\begin{figure}[t]
    \centering
    \includegraphics[width=1\linewidth]{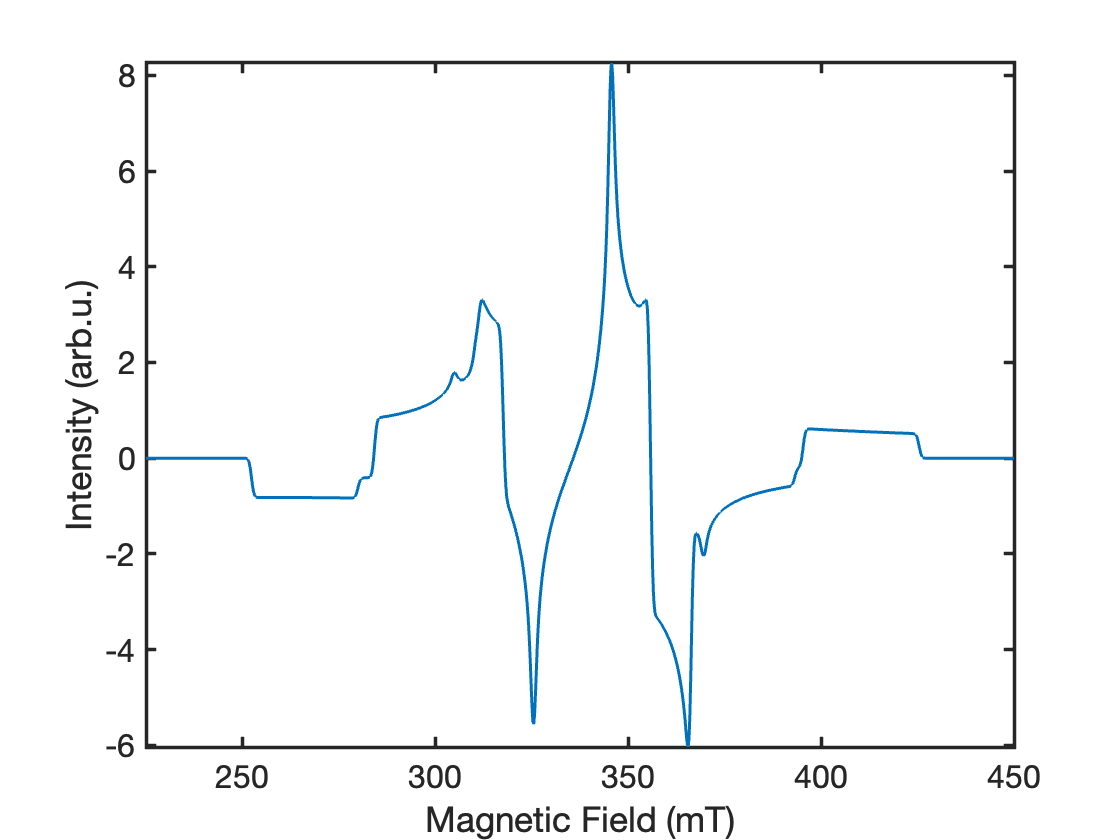}
    \caption{The simulated powder-average EPR spectrum at ca.\ 300 ns for singlet fission in lycopene dimers using the "pepper" function in EasySpin~\cite{StollES}. The polarisation pattern predicted is \textit{EAEAEA}. The intensity scale in presented in arbitrary units. The spectrometer frequency used in our simulation is $\nu = 9.5$ GHz.}
    \label{fig:EPR_Spec}
\end{figure}

\subsection{EPR simulations}

Following the results presented above, we now turn to the EPR simulation of our system pre-thermalisation. In order to obtain results which can be used in direct comparison with experimental evidence, we have used EasySpin~\cite{StollES,TaitES} to simulate the powder-average EPR spectra for a lycopene dimer in an H-aggregate. The original state of the program did not account for the biquadratic exchange term $J_2$ present in eqn.~\ref{eq:H_two_triplet}, which we have explicitly added into the program.

\begin{figure}[t]
    \centering
    \includegraphics[width=1\linewidth]{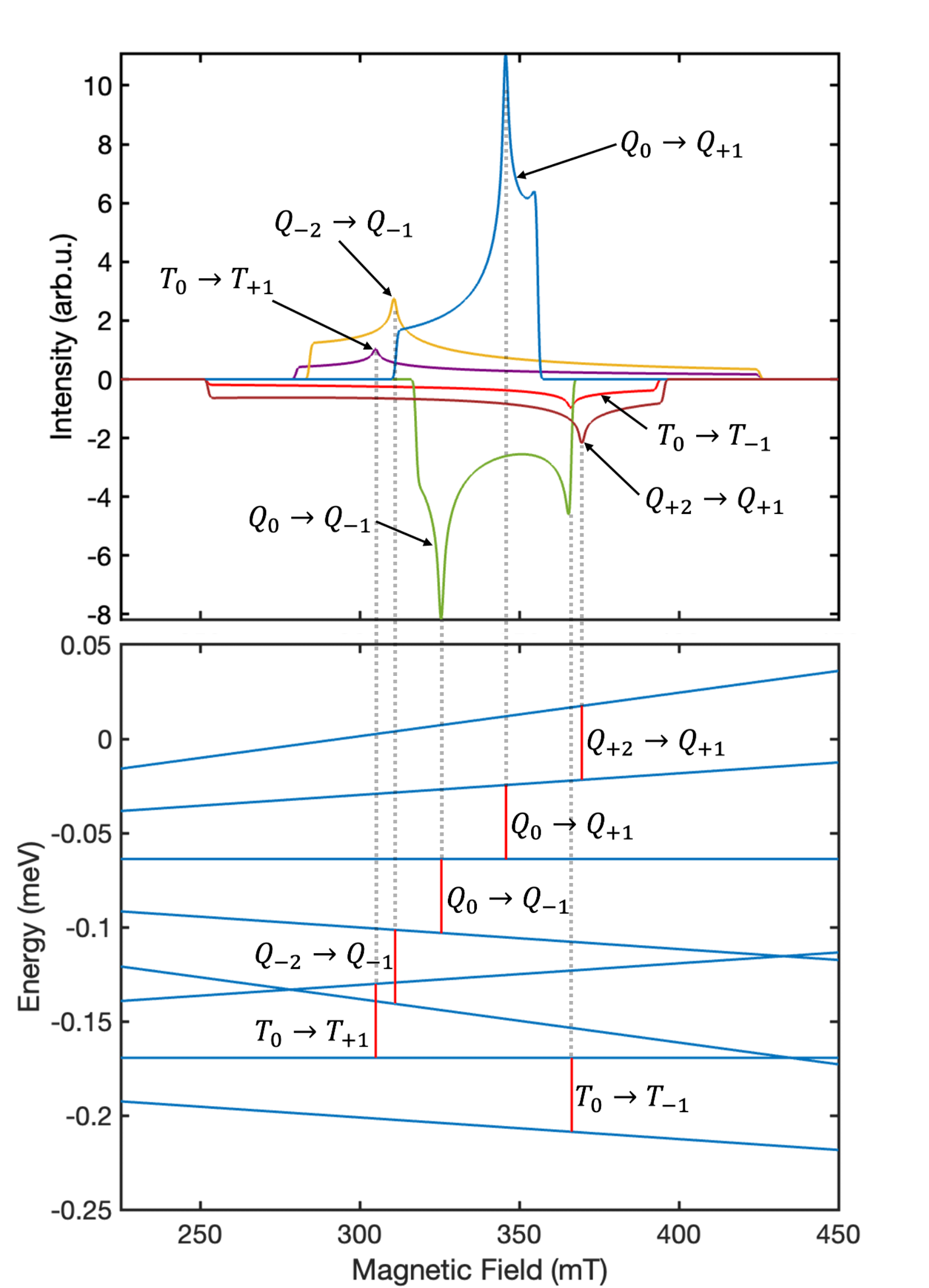}
    \caption{The six distinct EPR transitions illustrated in Fig.\ \ref{fig:EPR_Spec} are attributed to $\Delta M_S= \pm1$ transitions between the eigenstates of the system. As an example, the contribution arising from the transitions in the case where \textbf{B} is oriented along the principal axis \textbf{X} is shown. The energy scale was calculated using the "levelsplot" function in EasySpin~\cite{StollES} and only the energy differences are shown.
    The red vertical bars correspond to the resonance condition $h\nu = 0.039$ meV.}
    \label{fig:EPR_transitions}
\end{figure}

Fig.~\ref{fig:EPR_Spec} displays the simulated EPR spectrum of our system at ca.~300 ns, which corresponds to the experimental instrument response time~\cite{Weiss2017}. The contributing transitions are shown and labelled in Fig.~\ref{fig:EPR_transitions}. We identify six distinct $\Delta M_S = \pm1$ transitions:
\begin{itemize}
    \item four transitions between the high-spin quintet triplet-pair states, and
    \item two transitions between the high-spin triplet triplet-pair states.
\end{itemize}
We assume that there are no quintet $\Delta M_S = \pm 2$ transitions outside of the field range $225-450$ mT as a consequence of the strong exchange interaction overcoming the much weaker dipolar interaction that leads to mixing within the quintet spin sub-space. However, we also note that the intensity of such transitions can, in theory, provide a measure of the strength of the dipolar interaction within the triplet-pairs relative to that of the exchange coupling.

The overall spectrum is predicted to have an \textit{EAEAEA} (\textit{A} = absorption, \textit{E} = emission) absorption-emission pattern, which is in contrast to the frequently reported \textit{AEEAAE} pattern in experimental work on singlet fission in acene-type systems~\cite{Weiss2017, Tayebjee2017,Basel2017}. Recently, the existence of a coupled quintet triplet-pair intermediate has been unequivocally proved by EPR studies on singlet fission in acene-like systems~\cite{Weiss2017,Tayebjee2017,Basel2017}, which showed that the \textit{AEEAAE} polarisation can be obtained by a mixture of coupled quintet triplet-pairs and weakly coupled triplets.

\section{Conclusions}

We have presented a theory of singlet fission in carotenoid dimers as applied to lycopene dimers in an H-aggregate. The theory is based on the assumption that the identity of the intermediate state in longer chain carotenoids, $S^*$, is the "dark", intrachain exchange-bound triplet-pair state $1^1B_u^-$. We calculated the spectrum of the carotenoid dimer, where we showed that the exchange coupling is responsible for the large splitting between the spin subspaces, whereas the dipolar interaction can be treated as a perturbation that removes the degeneracies within the spin manifolds. 

The dynamical simulations were calculated using the quantum Liouville equation supplemented with two Lindblad dissipators corresponding to transverse ($T_2$) and longitudinal ($T_1$) spin-dephasing. The model predicts that at long times the effect of the spin-decoherence and thermalisation is to yield single, unentangled triplets on separate carotenoid chains, as shown by the scheme in Fig.\ \ref{fig:mechanism}. However, this does not mean that the triplets are noninteracting, which would require  triplet diffusion over the whole aggregate that is not possible in our model. This means that, even at long times, the triplets experience an exchange interaction. This is furthermore visible in our simulated EPR spectra, which display a distinct polarisation pattern caused by the residual intertriplet exchange interaction.

Whilst we have focussed our attention to the application of the theory to lycopene dimers starting from the $1^1B_u^-$ state, this model can be used to explain singlet fission in carotenoid aggregates for a diverse range of carotenoids. Similarly, there is no strict requirement to assume that the $1^1B_u^-$ state is the identity of the intermediate, meaning that other high-energy members of the $2A_g$--family can be responsible for the formation of the interchain singlet intermediate $^1|T...T\rangle_{1-2}$.

Further theoretical work should include the effect of triplet diffusion in the carotenoid aggregate. This can be accomplished by various means, notably either by an expansion of the basis set to explicitly include the new chains, or by introducing a spatial-dependence of the intertriplet exchange interaction. Similarly, real solids will posses conformational disorder in the packing of the molecules, so the effects of this disorder, expressed through the loss of the permutation symmetry of the dipolar interaction, should be explored in future work.


\section*{Conflicts of interest}
The authors have no conflicts to disclose.
\section*{Data availability}
See Supplementary Information
\section*{Acknowledgements}
We thank Claudia E.~Tait for the insightful discussion on EPR and EasySpin, and Eric R.~Bittner for discussion on the calculation of entanglement entropy.

\hspace{1cm}

\appendix{\large{\bf{Appendix A -- Perturbative Calculation of the Low-Energy Zero-Field Spectrum}}}\label{Appendix_A}

Our goal in this appendix is to determine the nine low-energy eigenstates, whose energy levels are shown  schematically in Fig.\ \ref{fig:ZF_Spec}.
We perform this calculation using the reduced two-triplet model, $\hat{H}_\textrm{reduced}$ (eqn.~\ref{eq:H_two_triplet}), whose  matrix representation is shown in the Supplementary Information. Due to the large difference in the magnitudes of the residual exchange interactions (i.e., $J_1$ and $J_2$) and the dipolar interactions (i.e., $D$ and $E$), we  treat the dipolar coupling as a perturbation that removes the degeneracies within the triplet and quintet spin-subspaces.

As mentioned in the main text, the dipolar coupling for a dimer in a perfect H-aggregate can only couple three (for $E\neq0$) of the quintet states to the singlet, leaving the remaining two quintet and three triplet states
with definite spin character. The effect of the dipolar interaction on the eigenvalues can be solved exactly for these triplets and two quintets. Within the reduced two-triplet model (eqn.~\ref{eq:H_two_triplet}) the eigenvalues are
\begin{eqnarray}
    &\textrm{E}_2^T =& -J_1 - J_2 - \frac{D}{3} - E,\\
    &\textrm{E}_3^T =& -J_1 - J_2 - \frac{D}{3} + E,\\
    &\textrm{E}_4^T =& -J_1 - J_2 + \frac{2D}{3},\\
    &\textrm{E}_6^Q =& J_1-J_2-\frac{D}{3}-E
\end{eqnarray}
and
\begin{eqnarray}
    &\textrm{E}_7^Q =& J_1-J_2-\frac{D}{3}+E.
\end{eqnarray}
 Taking the axis of spin-quantisation to be parallel to the principal \textbf{Z} axis, the corresponding spin-eigenstates are
\begin{eqnarray}
    &|\Psi_2\rangle =& \frac{1}{\sqrt{2}}\left(|1,+1\rangle -|1,-1\rangle\right),\\
    &|\Psi_3\rangle =& \frac{1}{\sqrt{2}}\left(|1,+1\rangle +|1,-1\rangle\right),\\
    &|\Psi_4\rangle =& |1,0\rangle,\\
    &|\Psi_6\rangle =& \frac{1}{\sqrt{2}}\left(|2,+1\rangle -|2,-1\rangle\right)
\end{eqnarray}
and
\begin{eqnarray}
    &|\Psi_7\rangle =& \frac{1}{\sqrt{2}}\left(|2,+1\rangle +|2,-1\rangle\right).
\end{eqnarray}

Before the perturbation, all of the quintet sublevels are degenerate. Since a linear combination of degenerate eigenstates is still an eigenstate with the same energy, the $|2,+2\rangle$ and $|2,-2\rangle$ states can mix to form the symmetry-adapted linear combinations
\begin{equation}
    ^{5}|\psi_\pm\rangle = \frac{1}{\sqrt{2}}\left( |2,+2\rangle \pm |2,-2\rangle \right).
\end{equation}
As a consequence of a symmetry property of $\textbf{H}_{\textrm{dipolar}}$ (see eqn.~(1) of the Supplementary Information), $^{5}|\psi_{-}\rangle$ is decoupled,
i.e., $|\Psi_8\rangle = {^5}|\psi_-\rangle$, with an exact energy
\begin{equation}
    \textrm{E}_8^Q = J_1-J_2+\frac{2D}{3}.
\end{equation}

The remaining three states that need consideration are $|0,0\rangle$, $|2,0\rangle$ and ${^5}|\psi_+\rangle$. The matrix representation of the perturbation in the basis of these three states is
\begin{equation}
    \textbf{H}_{\textrm{pert}} =
    \left[
    {
    \begin{array}{ccc}
        0 & \frac{\sqrt{8}D}{3} & \frac{4E}{\sqrt{6}}\\
        \frac{\sqrt{8}D}{3} & -\frac{2D}{3} & \frac{2E}{\sqrt{3}}\\
        \frac{4E}{\sqrt{6}} & \frac{2E}{\sqrt{3}} & \frac{2D}{3}\\
    \end{array}
    }
    \right].
    \label{eq:pert_matrix1}
\end{equation}

Again, before the perturbation the degenerate states $|2,0\rangle$ and $^{5}|\psi_+\rangle$ mix to form the linear combinations that will undergo the least change under the perturbation, $^{5}|\phi_\pm\rangle$. Within the basis \{$|0,0\rangle$, $^{5}|\phi_+\rangle$, $^{5}|\phi_-\rangle$\}, where the states ${^5}|\phi_\pm\rangle$ are
\begin{eqnarray}
    &{^5}|\phi_+\rangle \approx& {^5}|\psi_+\rangle + \sqrt{\frac{3E^2}{4D^2}}|2,0\rangle
\end{eqnarray}
and
\begin{eqnarray}
    &{^5}|\phi_-\rangle \approx& |2,0\rangle - \sqrt{\frac{3E^2}{4D^2}}{^5}|\psi_+\rangle,
\end{eqnarray}

the perturbation is
\begin{equation}
    \textbf{H}_{\textrm{pert}} =
    \left[
    {
    \begin{array}{ccc}
        0 & \sqrt{6}E & \frac{\sqrt{8}}{3}D \\
        \sqrt{6}E & \frac{2}{3}\sqrt{D^2+3E^2} & 0\\
        \frac{\sqrt{8}}{3}D & 0 & -\frac{2}{3}\sqrt{D^2+3E^2}\\
    \end{array}
    }
    \right].
    \label{eq:pertsmall3}
\end{equation}

It follows that the remaining eigenvalues are approximately
\begin{eqnarray}
    &\textrm{E}_1^S \approx&-2J_1-4J_2-\frac{1}{\Delta E}\left( \frac{8D^2}{9}+6E^2 \right),\\
    &\textrm{E}_5^Q \approx&J_1-J_2-\frac{2}{3}\sqrt{D^2+3E^2}+\frac{8D^2}{9\Delta E}
\end{eqnarray}
and
\begin{eqnarray}
    &\textrm{E}_9^Q \approx&J_1-J_2+\frac{2}{3}\sqrt{D^2+3E^2}+\frac{6E^2}{\Delta E},
\end{eqnarray}
where  $\Delta E = 3\left(J_1+J_2\right)$.
Here  the superscripts \textit{S} and \textit{Q} refer to the spin manifold to which the corresponding eigenstate can be best attributed.
 As we expect from a qualitative assessment of the mixing between nondegenerate states, the perturbation stabilises the lowest energy eigenstate whilst destabilising the highest energy eigenstate. The corresponding eigenstates are
\begin{eqnarray}
    &|\Psi_1\rangle \approx& |0,0\rangle -\frac{\sqrt{6}E}{\Delta E}|\phi_+\rangle - \frac{\sqrt{8}D}{3 \Delta E}|\phi_-\rangle,\\
    &|\Psi_5\rangle \approx& |\phi_-\rangle + \frac{\sqrt{8}D}{3 \Delta E}|0,0\rangle
\end{eqnarray}
and
\begin{eqnarray}
    &|\Psi_9\rangle \approx& |\phi_+\rangle + \frac{\sqrt{6}E}{\Delta E}|0,0\rangle.
\end{eqnarray}
Therefore, eigenstate $|\Psi_1\rangle$ has mostly singlet character, $|\Psi_5\rangle$ has mostly $|2,0\rangle$ character and $|\Psi_9\rangle$ is a linear combination of mainly $|2,\pm2\rangle$ with some singlet and $|2,0\rangle$ character.

\hspace{1cm}

\appendix{\large{\bf{Appendix B -- Measurement of Entanglement}}}\label{Appendix_B}

\begin{figure}
    \centering
    \includegraphics[width=1\linewidth]{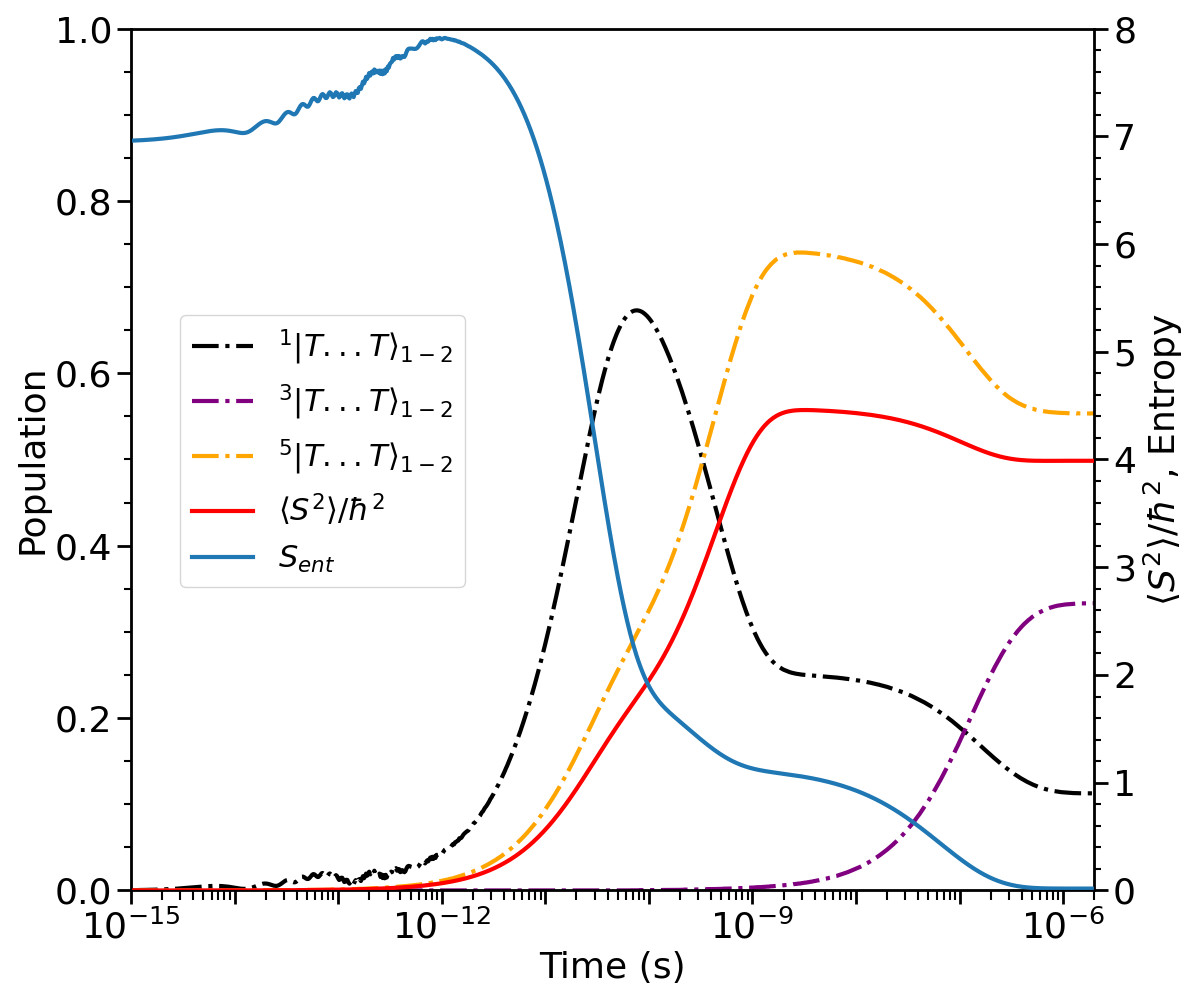}
    \caption{The entanglement entropy, $S_{ent}$, and expectation value of the spin, $\langle S^2 \rangle$.
    The populations of the interchain triplet-pair states are also shown.
    Once the system reaches thermal equilibrium at ca.~2 $\mu$s the values of the entropy and $\langle S^2\rangle$ are approximately 0 and $4\hbar^2$, respectively, corresponding to complete singlet fission to a pair of unentangled triplets. We see that this is equivalent to equal populations of the triplet-pair states, i.e., $P\left({^1}|T...T\rangle_{1-2}\right)=\frac{1}{9}$, $P\left({^3}|T...T\rangle_{1-2}\right)=\frac{3}{9}$ and $P\left({^5}|T...T\rangle_{1-2}\right)=\frac{5}{9}$.}
    \label{fig:S_ent}
\end{figure}

The population dynamics provides us with evidence to suggest that the triplet-pair eigenstates have been effectively thermally mixed to result in two spin-uncorrelated, single triplets on separate chains in the long time limit. This is equivalent to complete singlet fission in the carotenoid dimer. Here, we calculate a more direct measure of the entanglement of the triplet-pair, namely the von Neumann entropy, defined as
\begin{equation}
    S_{i}=-\sum_{\alpha}\omega_\alpha~\textrm{log}_2(\omega_\alpha),
\end{equation}
where $S_i$ is the entropy of the (sub)system \textit{i} and $\{ \omega \}$ are the eigenvalues of the time-dependent density operator.
The entanglement entropy is defined as
\begin{equation}
    S_{ent} = S_{A} + S_{B} - S_{AB},
\end{equation}
where $A$ and $B$ denote the two triplets. The reduced density matrix for one of the subsystems (e.g., $A$) is found by tracing over the degrees of freedom of the other subsystem,
\begin{equation}
  \hat{\rho}_A = \textrm{Tr}_B\left(\hat{\rho}\right),
\end{equation}
or equivalently written in a form which can be of more use computationally,
\begin{equation}
  \rho^{A}_{ij}=\sum_{k\in B}\rho_{(ik)(jk)}.
\end{equation}

The entanglement entropy and the spin expectation value as functions of time are illustrated in Fig.~\ref{fig:S_ent}. We notice that the entanglement entropy increases from its initial value during the coherent population transfer between the intrachain states via the singlet interchain state, followed by a monotonic decrease during the population transfer between the interchain states. At thermal equilibrium, $S_{ent}\approx 0$ which further explicitly displays the complete de-entanglement of the triplets. In addition, the expectation value of the spin is that of two triplets, $\langle S^2 \rangle \approx 4 \hbar^2$, with the populations of the singlet, triplet and quintet spin-subspaces reaching $\frac{1}{9}$, $\frac{3}{9}$ and $\frac{5}{9}$, respectively. We therefore conclude that complete singlet fission has taken place at ca.~2 $\mu$s.



\balance
\vspace{0.5cm}

\renewcommand\refname{Notes and references}

\bibliography{References} 
\bibliographystyle{rsc} 

\end{document}